\documentclass[twocolumn,english,floatfix]{revtex4-1}
\usepackage[T1]{fontenc}
\usepackage[latin9]{inputenc}
\usepackage{color}
\usepackage{amsmath}
\usepackage{graphicx}
\usepackage{braket}
\usepackage{bbold}
\usepackage{natbib}

\usepackage[colorlinks=true,urlcolor=blue,citecolor=blue,linkcolor=blue]{hyperref}

\makeatother

\usepackage{babel}

\begin{document}
	
	\title{Leakage reduction in fast superconducting qubit gates via optimal control}
	
	\author{M. Werninghaus$^1$, D. J. Egger$^1$, F. Roy$^2$, S. Machnes$^2$, F. K. Wilhelm$^2$, S. Filipp$^1$}
	\affiliation{$^1$IBM Research GmbH, Zurich Research Laboratory, S\"aumerstrasse 4, 8803 R\"uschlikon, Switzerland\\
		$^2$Theoretical  Physics,  Saarland  University,  66123  Saarbr\"ucken,  Germany}
	
	\begin{abstract}
		Reaching high speed, high fidelity qubit operations requires precise control over the shape of the underlying pulses. For weakly anharmonic systems, such as superconducting transmon qubits, short gates lead to leakage to states outside of the computational subspace. Control pulses designed with open-loop optimal control may reduce such leakage. However, model inaccuracies can severely limit the usability of such pulses. We implemented a closed-loop optimization that simultaneously adapts all control parameters based on measurements of a cost function built from Clifford gates. By parameterizing pulses with a piecewise-constant representation that matches the capabilities of the control hardware we create a $4.16~\rm{ns}$ single-qubit pulse with $99.76\,\%$ fidelity and $0.044\,\%$ leakage. This is a seven-fold reduction of the leakage rate of the best DRAG pulse we have calibrated at such short durations on the same system.

	\end{abstract}
	
	\date{\today}
	
	\maketitle
	
\section{Introduction}

Superconducting qubits are a promising candidate to realize large scale quantum computing systems \cite{Devoret2013,Krantz2019,Wendin2017}.
The architecture is scalable \cite{Acin2018}, microwave control electronics are well developed and readily available, and transmon-type qubit designs \cite{Koch_2007} allow for stable operations.
To accurately manipulate the quantum state of the weakly anharmonic qubit, control methods have been steadily improved to address common problems such as frequency crowding \cite{Motzoi2009, Schutjens2013,Vesterinen2014} and cross talk \cite{Sheldon2016b}. In particular, with the powerful tools provided by open-loop optimal control theory preparing target states \cite{Rojan2014,Reich_2013,Malis2019} and gates \cite{Sporl2007,Egger2014} can be realized with high fidelity.
In these methods numerically simulated system models are used to optimize hundreds of parameters that determine the shape of the control fields applied to the quantum system \cite{Glaser2015,Machnes2018}.
When the system model is accurate enough, the optimized control pulses can immediately be applied in experiment, yielding high performance and reliable control \cite{Khaneja2005,Lapert,Heeres2017}.
%An alternative approach is to perform optimal control in a closed-loop search directly on the experimental system \cite{Kelly_2014}. %TODO move this sentence 
However, applying such control methods to superconducting qubits produces less accurate results in comparison to ion traps \cite{Timoney2008,Mount2015} and nuclear magnetic resonance systems \cite{Glaser2015,Kobzar} since models with sufficient accuracy are not available for superconducting qubits. Effects that are hard to accurately reproduce in simulation include instrument noise, transfer functions \cite{Rol2019a,Jerger2019,Gustavsson2013}, additional modes and coupling to unwanted quantum systems \cite{Muller2015}. As a result, pulse shaping for superconducting qubits requires closed-loop optimal control, i.e. direct optimization on the experimental system, which limits the amount of tunable parameters defining the pulse shapes \cite{Egger2014, Kelly_2014}.

Furthermore, optimal control can suppress leakage out of the computational subspace occurring for fast qubit gates. Fast gates are required to lower the limits on gate errors set by decoherence. Moreover, in combination with short readout times \cite{Heinsoo2018} and fast qubit reset \cite{Egger2018c,Magnard2018a} fast gates are useful to reduce the overall execution time of quantum algorithms, such as variational quantum eigensolvers \cite{Moll2018} that require many repetitions to gather statistics on quantum measurements \cite{Torlai2019,Ganzhorn2019a,Wecker2015a}. However, fast qubit gates suffer from leakage effects and additional unitary errors caused by the large bandwidth of the short control pulse \cite{Motzoi2009}. Reducing leakage out of the computational subspace is paramount for error correction since correcting such errors requires significantly more resources than correcting errors in the computational sub-space \cite{Fowler2013, Suchara2015, Wood2018a}.

In this work, we increase the fidelity of short duration single-qubit gates.
Control shapes are optimized in a closed-loop fashion to capture the full system dynamics and avoid model limitations.
The resulting pulses significantly mitigate leakage effects.
As optimization is limited by the experimental runtime, we investigate the time budget of the setup. We employ methods such as restless measurements \cite{Rol2017} to speed-up the optimization and use the CMA-ES algorithm \cite{Hansen} instead of, for instance, Nelder-Mead \cite{Nelder1965a} to handle the large number of parameters.
With these improvements we experimentally optimize pulses with up to 55 parameters.

\section{RESULTS}
\begin{figure}
	\includegraphics[width=.95\columnwidth]{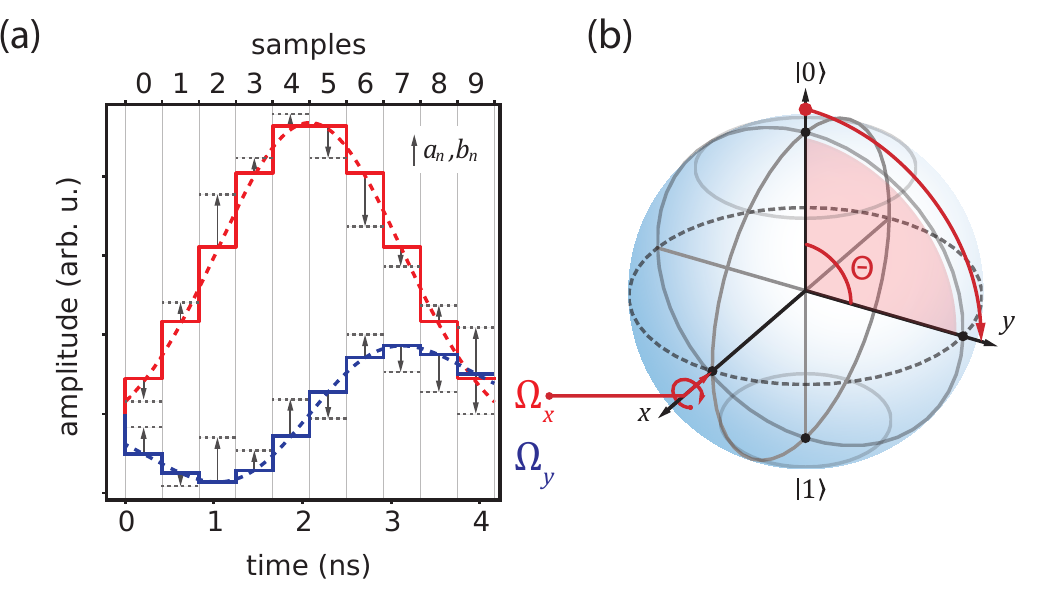}
	\caption{\label{fig:pulse} 
		(a) The dashed lines show the analytic DRAG pulse, with $\Omega_x$ in red and $\Omega_y$ in blue. 
		The solid lines show the same pulse sampled by the AWG. The optimization parameters $a_n$ and $b_n$ of the piecewise-constant pulse are depicted as modifications of the sampled DRAG pulse by grey arrows and dashed lines.
		(b) Ideal Bloch sphere trajectory of the  $\Omega_x$-pulse.
		The rotation angle $\Theta$ is given by the total area under the pulse.}
\end{figure}
The system consists of two transmon-type fixed-frequency superconducting qubits \cite{Koch2007a} coupled by a flux tunable coupler \cite{McKay2016,Roth2017}.
Experiments are carried out on one of the qubits with a transition frequency of $\omega_{01}/2\pi = 5117.22~\rm{MHz}$, an anharmonicity $\Delta/2\pi = -315.28~\rm{MHz}$ and coherence times of  $105~\mu\rm{s}$ and $39~\mu\rm{s}$ for $T_1$ and $T_2$, respectively. 
The qubit is controlled by microwave pulses applied via a readout resonator capacitively coupled to the qubit.
Pulses consist of two control components $\Omega_x(t)$ and $\Omega_y(t)$, which are combined into a single drive signal 	$\Omega=\Omega_x+i\Omega_y$.
The pulse is up-converted to the qubit frequency using a microwave vector signal generator in a single-sideband configuration. 

The in-phase- and quadrature-components in the mixing process are the real and imaginary parts of the pulse modulated at an intermediate sideband frequency $\Omega(t)\exp\{i(\omega_\text{ssb}t+\phi)\}$ with $\omega_\text{ssb}/2\pi = 100~\rm{MHz}$. Synthesizing this signals by an arbitrary waveform generator (AWG) results in real-time control over phase, frequency and amplitude \cite{SSB_whitepaper}.  

In a frame rotating at the qubit frequency, the transmon Hamiltonian is given by
\begin{equation}
\label{eq:hamiltonian}
\frac{\hat{H}^R}{\hbar} = \Delta \ket{2}\!\!\bra{2} +\frac{\Omega_x(t)}{2}\sum_{j=1}^2\hat{\sigma}^x_{j,j-1}+\frac{\Omega_y(t)}{2}\sum_{j=1}^2\hat{\sigma}^y_{j,j-1},
\end{equation} 
where terms rotating at twice of the qubit frequency have been omitted.
The $i^\text{th}$ level of the transmon is denoted by $\ket{i}$.
%Therefore, the parametrization of the envelope shape is limited by the physical capabilities of the AWGs such as sampling rate and 16-bit amplitude resolution.
The operators $\hat\sigma^x_{j,j-1} = \sqrt{j}\left(\ket{j}\!\!\bra{j-1}+\ket{j-1}\!\!\bra{j}\right)$ and $\hat\sigma^y_{j,j-1} = i\sqrt{j}\left(\ket{j}\!\!\bra{j-1}-\ket{j-1}\!\!\bra{j}\right)$ couple adjacent energy levels. 
Therefore, $\Omega_{x}$-pulses at the resonance frequency $\omega_{01}$ drive rotations about the $x-$axis of the Bloch sphere spanned by $\{\ket{0},\ket{1}\}$, see Fig.~\ref{fig:pulse}. The total area of the pulse envelope defines the rotation angle  $\Theta$. 
The rotation axis can be freely chosen in the $xy$-plane by changing the phase of the drive signal $\phi$. By selecting $\phi=n\pi/2$ ($n=0,1,\ldots$) and  $\Theta = \pi/2$, $\pm X/2$ and $\pm Y/2$ single-qubit operations are realized.

Since transmons have a low anharmonicity, fast pulses with a wide frequency response lead to leakage out of the computational subspace defined by the two lowest-lying energy eigenstates.
This process is suppressed by derivative removal gates (DRAG) \cite{Motzoi2009,Chen2016b,Sheldon2016}, designed to reduce leakage and phase errors caused by inadvertent driving of the $\ket{1}\leftrightarrow\ket{2}$ transition.
The first-order DRAG correction (Fig.~\ref{fig:pulse}(a); dashed lines) to a Gaussian shaped pulse $\Omega_x(t) = A\exp\left\{-t^2/(2\sigma^2)\right\}$ with amplitude $A$ and width $\sigma$, is
\begin{equation}
\label{eq:DRAG}
\Omega_\text{DRAG}(t)= \Omega_x(t)+i\frac{\beta}{\Delta}\frac{d\,\Omega_x(t)}{dt}.
\end{equation} 
The correction in the imaginary component of $\Omega_\text{DRAG}(t)$ with the scaling parameter $\beta$ eliminates the spectral weight of the pulse at the $\ket{1}\leftrightarrow\ket{2}$ transition.

Although being designed for fast, short gates DRAG fails to produce high fidelities when the gate duration is lower than $\sim10/\Delta$ \cite{Motzoi2009}. To overcome this, either higher-order correction terms or pulses with more degrees of freedom have to be employed. To find suitable pulses we use a parameterization that applies a correction $\delta_n=a_n+ib_n$ at each point in time to a calibrated DRAG pulse, similar to common optimal control approaches \cite{Schulte-Herbruggen2012,Khaneja2005}. This results in a list of piecewise-constant control amplitudes 
\begin{equation} \label{eq:pwc}
\Omega_n =\Omega_\text{DRAG}(n\Delta t) + \delta_n,
\end{equation}
as shown in Fig.~\ref{fig:pulse}(a). The time discretization $\Delta t$ is naturally given by the sampling rate of the AWG generating the pulse envelope. We use a Zurich Instruments HDAWG \cite{ZI} operating at a sampling rate of $f_s = 2.4~\rm{GS/s}$. The optimization parameters are the amplitude corrections $a_n$ and $b_n$ to the $n$-th sample of $\Omega_x$ and $\Omega_y$, respectively, with the initial guess $a_n=b_n=0$. 

\subsection{Pulse parameter optimization}

\begin{figure}
	\includegraphics[width=1.0\columnwidth]{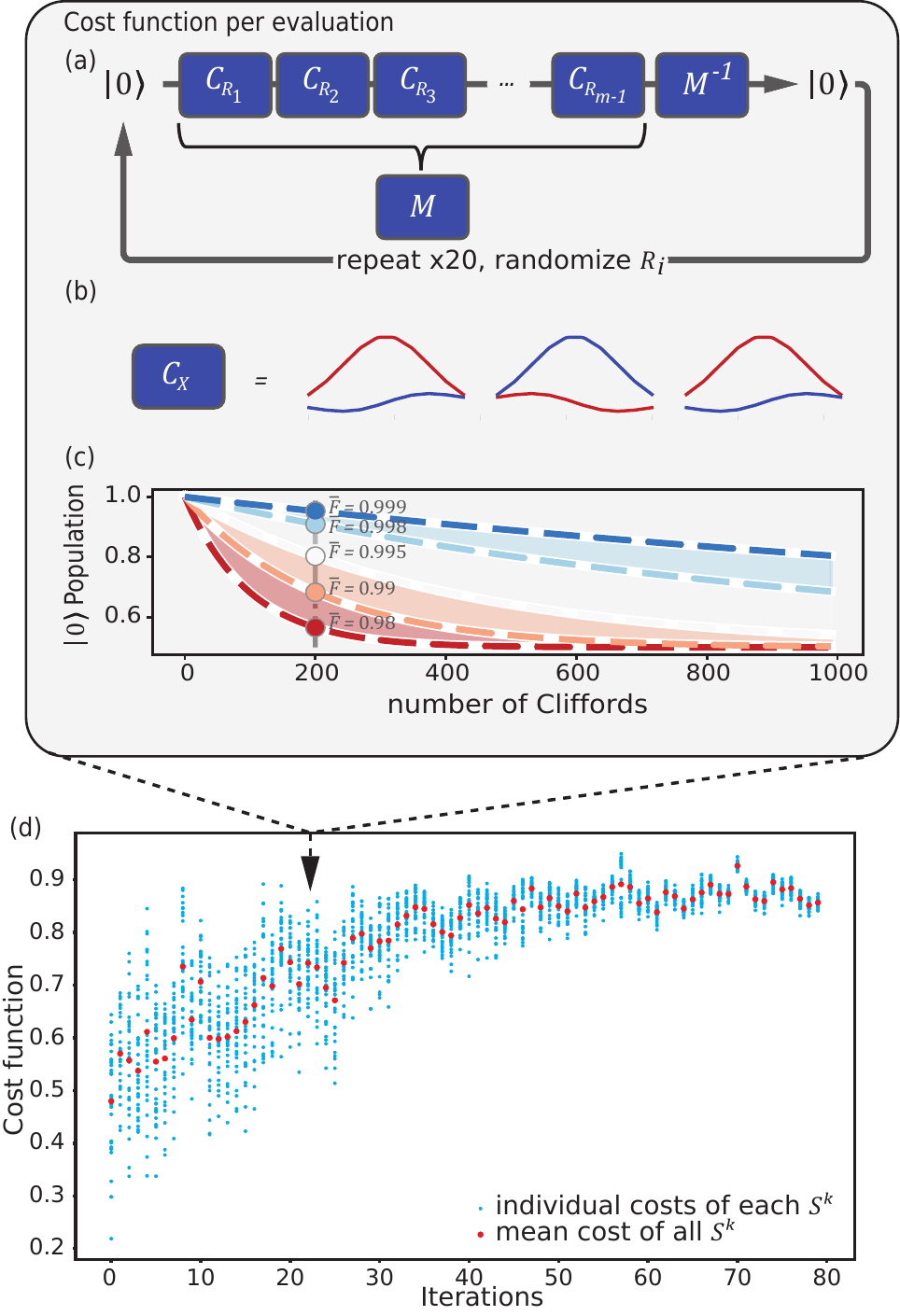}
	\caption{\label{fig:optimizer} 	(a) Single-qubit Clifford gate sequence of length $m$.
		(b) Schematic visualization of the composition of a Clifford gate from $\pm X/2, \pm Y/2$ pulses based on a specific pulse shape.
		The $\Omega_x$ and $\Omega_y$ components are displayed in red and blue, respectively.
		(c) Simulated datasets showing the cost function for $m=120$ Clifford gates as a point on the full randomized benchmarking curves for several fidelities.
		(d) Experimental data of a full optimization run for a 23 dimensional parameter space.
		The blue points represent the cost function of each candidate pulse shape based on a unique parameter set $\mathcal{S}^k$ evaluated using 20 Clifford sequences. The red points represent the average cost function at each iteration of the optimizer.}
\end{figure}

Since the parametrization in Eq.~(\ref{eq:pwc}) no longer permits an individual optimization of each parameter we simultaneously optimize all of them using the Covariance Matrix Adaptation - Evolution Strategy (CMA-ES) optimization algorithm \cite{Hansen} (see Methods section). It is based on generating sets of parameters $\mathcal{S}^k$ that describe $k=1,...,\lambda$ different pulse shapes as candidate solutions. The parameters in $\mathcal{S}^k$ are defined by the parametrization of the pulse shape. The fidelity of each candidate solution is evaluated by a cost function, which serves to generate a new set of candidate solutions. This process is repeated until convergence is reached and the best solution is found.

As a cost function we use randomized benchmarking (RB) sequences with a fixed number of $m$ Clifford gates \cite{Kelly_2014} averaged over $K$ sequence realizations, see Fig.~\ref{fig:optimizer}(a). This corresponds to evaluating only a single point in a standard RB measurement \cite{Magesan_SRB_2012,Magesan2011a} which reduces the runtime to evaluate the cost function. We construct the Clifford gates by composing $\pm X/2$ and $\pm Y/2$ pulses, each based on the pulse shape defined by $\mathcal{S}^k$, see Fig.~\ref{fig:optimizer}(b). The average ground state population $\overline{p}_0(m)$ of the final qubit state defines the cost function, which is maximized by the optimizer. 
To estimate the fidelity of the optimized pulses we finally perform a full randomized benchmarking measurement.

\subsection{Fidelity estimates of optimized short pulses}
\begin{figure}
	\includegraphics[width=0.95\columnwidth]{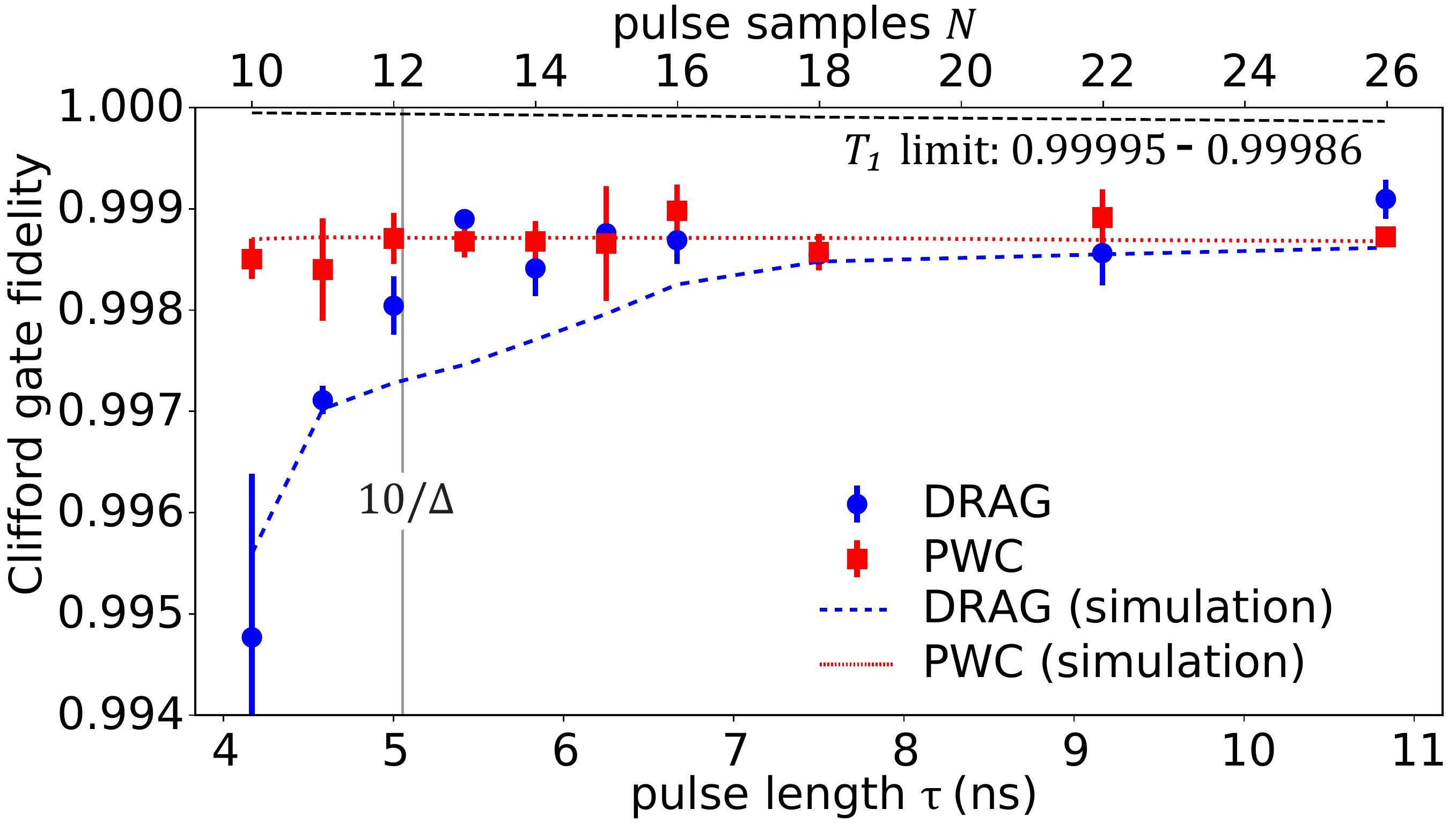}
	\caption{\label{fig:results} Fidelity measured with RB as a function of pulse length for optimized DRAG (blue circles) and piecewise-constant pulses (red squares). Simulated fidelities are shown with red dashed and blue dotted lines (see Methods). The dashed black line indicates the $T_1$ limit on the gate fidelity.}% estimated as $\exp\{-N\Delta t/T_1\}$ where $N\Delta t$ is the gate duration.}
\end{figure}
We optimize single-qubit pulses of varying duration ranging from $N=10$ to $N=26$ samples per pulse, corresponding to a duration $\tau=N\cdot f_s$ ranging from $4.16~\rm{ns}$ to $10.83~\rm{ns}$. We use  $K=20$ sequences of $m=120$ Clifford gates. Each sequence is measured 1000 times using the restless measurement protocol \cite{Rol2017} at a rate of $100~\rm{kHz}$. We first use the CMA-ES based optimization procedure to calibrate DRAG pulses, defined in Eq.~(\ref{eq:DRAG}). For this we choose the amplitude $A$, the DRAG parameter $\beta$ and the sideband frequency $\omega_{\rm ssb}$ as optimization parameters, i.e. $\mathcal{S} = \left\lbrace A, \beta, \omega_{\rm ssb}\right\rbrace$. The results of our CMA-ES based calibration is shown in Fig.~\ref{fig:results} (blue circles). The resulting fidelities compare well with standard sequential error amplification calibration methods \cite{Sheldon2016}. The optimized DRAG pulse then serves as initial guess for a second optimization step in which we extend $\mathcal{S}$ by the amplitude corrections to $\mathcal{S}' = \left\lbrace A, \beta, \omega_{\rm ssb},a_1,b_1,...,a_N,b_N\right\rbrace$.

For gates longer than $\tau = 6~\rm{ns}$ we find a constant fidelity of $F = 99.87(1)\,\%$ both for the DRAG pulse and the piecewise-constant optimized pulse, see Fig.~\ref{fig:results}. 
For gates shorter than $6~\rm{ns}$ we observe a decrease of fidelity for the DRAG pulses consistent with the $10/\Delta$ limit (see the grey line in Fig.~\ref{fig:results}), while the fidelity of the piecewise-constant optimized pulses remains constant even for the shortest gate duration.
Drive power limitations prevent us from implementing gates shorter than $4~\rm{ns}$.

\begin{figure}
	\includegraphics[width=0.95\columnwidth]{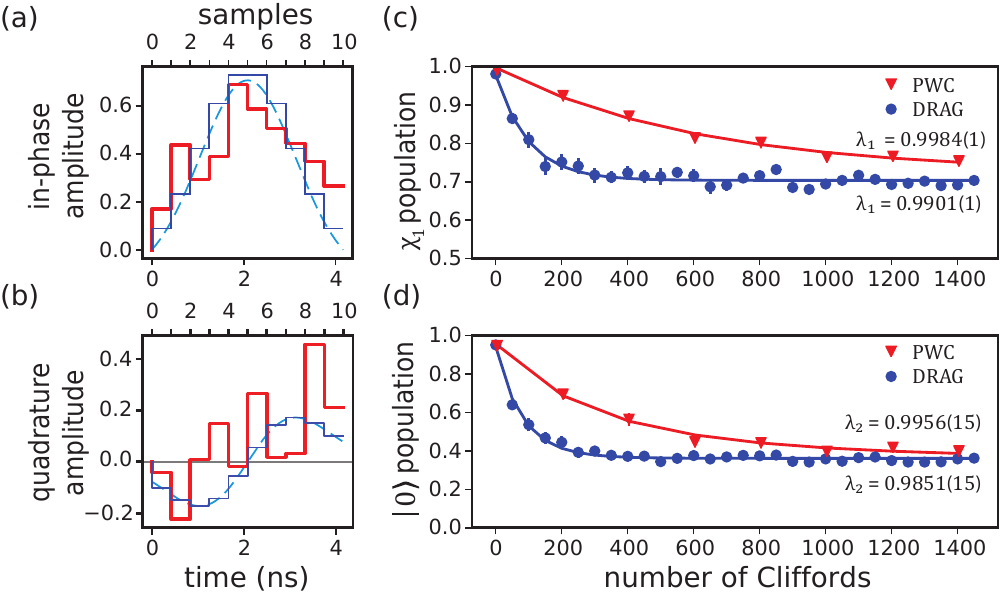}
	\caption{\label{fig:results_pulse}(a) In-phase and (b) quadrature amplitude component of the pulse envelope before (blue) and after the piecewise-constant optimization (red), as represented in AWG memory. (c) Remaining population in the computational subspace $\chi_1$ for randomized benchmarking measurements using pulses based on the DRAG and optimized piecewise-constant pulses. The decay constant $\lambda_1$ characterizes the population remaining in $\chi_1$. (d) Full leakage RB analysis characterization using a double decay with decay constants $\lambda_1$ and $\lambda_2$ for leakage and standard errors, respectively.}
\end{figure}

To assess the influence of leakage on the shortest $4.16~\rm{ns}$ pulse displayed in Fig.~\ref{fig:results_pulse}(a) and Fig.~\ref{fig:results_pulse}(b) we follow the leakage randomized benchmarking protocol outlined in \cite{Wood2018a}.
The leakage RB analysis requires measuring the probabilities $p_j$ to occupy the states $\ket{j}$ with $j\in\{0,1,2\}$ after the standard RB gate sequences.
The probability $p_{\chi_1} = p_0+p_1=1-p_2$ of remaining in the computational subspace $\chi_1 = \lbrace\ket{0},\ket{1}\rbrace$ is fitted using the decay model $A+B\lambda_1^n$ to find the average leakage per Clifford $L_1 = \left(1-A\right)\left(1-\lambda_1\right)$.
Here $n$ is the number of Clifford gates while $A$, $B$, and $\lambda_1$ are fit parameters.

Using the extracted leakage decay $B\lambda_1^n$ we fit $p_0(n)$ using the double decay model $A_0+B\lambda_1^n+C_0\lambda_2^n$ to find the average Clifford gate fidelity

\begin{equation}
\label{eq:fidelity}
\overline{F}=\frac{1}{2}\left[\lambda_2+1-L_1\right].
\end{equation}
The leakage rate of the optimized piecewise-constant pulses $L_1^{\text{PWC}}=0.044(25)\,\%$ is five times lower than the leakage rate of the DRAG pulse $L_1^{\text{DRAG}}=0.29(3)\,\%$, see Fig.~\ref{fig:results_pulse}(c).
Additionally, we observe a reduction of standard errors from $1-\lambda_2^\text{DRAG}=1.49(15)\,\%$ to $1-\lambda_2^\text{PWC}=0.44(15)\,\%$, see Fig.~\ref{fig:results_pulse}(d).
The resulting average fidelity per Clifford gate, computed using Eq.~(\ref{eq:fidelity}), is $\overline{F}_\text{PWC}=99.76(8)\,\%$ for the piecewise-constant pulse and $\overline{F}_\text{DRAG}=99.11(8)\,\%$ for the DRAG pulse.

\section{DISCUSSION}

Our results show that optimal pulse shaping using a piecewise-constant basis improves the gate fidelity of short pulses, reducing leakage errors by a factor of seven and standard errors by a factor of three.
At longer gate durations, controlling the pulse shapes beyond analytical DRAG pulses does not improve the fidelity.
All our pulses, aside from the DRAG pulses shorter than $5.5~\rm{ns}$, are limited to an error per gate of $0.13(1)\,\%$ on average.

%We therefore attribute the remaining error to effects unrelated to the pulse shape.
The fidelities that we measured are, however, not limited by the $T_1$-time, which sets an error per gate limit of $5\cdot10^{-5}$, see Fig.~\ref{fig:results}. Instead, the fidelity limitation we observe may be explained by a dephasing proportional to the Rabi rate of the drive \cite{Sheldon2016}, as illustrated by the simulated fidelities shown in Fig.~\ref{fig:results} (see Methods).

The improvements with more complex pulse shapes come at the expense of long calibration times.
Optimizing the longest pulse shape with $N=26$ samples (i.e. $55$ parameters) took up to 25 hours.
To understand how this time can be reduced we have measured the time taken to create the pulse sequences, initialize the control electronics, and gather the data (see Methods section).
Creating the pulse sequences and initializing the control electronics at each iteration consumes the most time.
Gathering the required data is only a small fraction of the total experimental run time.
With further improvements of the control electronics, for instance an internal generation of the $100~\rm{MHz}$ side-band modulation, we expect further significant reductions in the overall runtime of the optimizer.

Our work demonstrates that optimizing -- or calibrating -- pulses with up to 55 parameters is experimentally feasible.
This opens up the possibility to explore more elaborate optimal control methods on superconducting qubit platforms.  
We plan to extend this scheme to multi-qubit gates, where system dynamics are more complex and analytic optimal control methods are not as developed as for single-qubit gates \cite{Machnes2018}.
While a piecewise-constant parametrization, as done for single-qubit gates, is harder due to the long duration of two-qubit gates, other analytical pulse representations, such as its spectral components, will be explored to improve on gate performance.

\section{METHODS}

\label{sec:optimizer}
\begin{figure}
	\centering
	\includegraphics[width=.95\columnwidth]{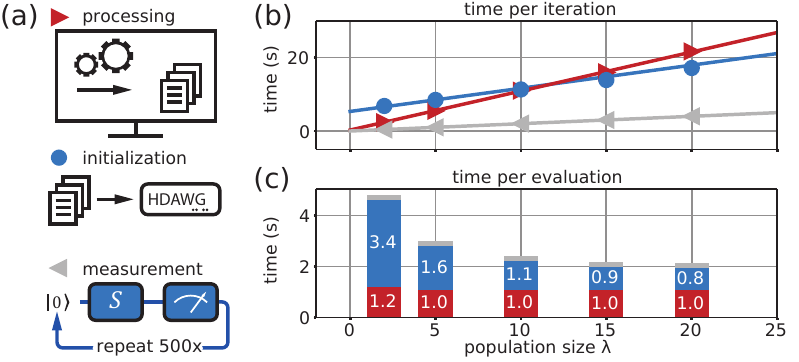}
	\caption[width=1\textwidth]{(a) Experimental runtime consisting of processing the pulse sequences (red right triangles), initializing the setup (blue circles) and measuring the cost function (grey left triangles).
		(b) Time per iteration of CMA-ES split into those three categories. In one iteration the cost function of each candidate solution in the whole population of size $\lambda$ is measured.
		Error bars are smaller than the size of the data points.
		(c) Time per \textit{evaluation}, as a function of population size. Each candidate solution in a given population requires one evaluation.
		As the population size increase the experimental run-time to evaluate a full iteration increases and the average time to evaluate a candidate solution decreases.  
	}
	\label{fig:time}
	
\end{figure}

To optimize all parameters of the pulse shape simultaneously on the experimental setup, we have chosen the Covariance Matrix Adaptation - Evolution Strategy (CMA-ES) optimization algorithm as a noise-resilient and time-efficient optimizer \cite{Hansen}.
This algorithm optimizes a population of $\lambda$ candidate solutions which are normally distributed in the parameter space. The center and spread of the distibution are chosen as starting conditions of the optimization.

Generally, the choice of the population size $\lambda$ is a trade-off between fast convergence speed and avoiding local optima \cite{Hansen}.
However, experimentally we have to consider the time required to process the pulse sequences (i.e. the time required to compile the pulse sequences into AWG files), to initialize the hardware (including data transfer) and to measure the cost function for different population sizes $\lambda$, see Fig.~\ref{fig:time}.
We benchmark these three times using a set of 20 Clifford gate sequences  per candidate solution, each with $100$ Clifford gates. 
By dividing the total time required to evaluate the entire population by $\lambda$ we calculate the effective time required to asses a single candidate solution.
This allow us to gauge how efficiently the hardware is being used, see Fig.~\ref{fig:time}(c).
The instrument initialization introduces a constant overhead, see Fig.~\ref{fig:time}(a), which decreases the efficiency of the optimization algorithm for lower population sizes.
Single-point optimizers, such as Nelder-Mead \cite{Nelder1965a}, are thus an inefficient choice.
However, when evaluating larger populations the contribution of the constant offset of the initialization is distributed over multiple measurements, leading to a convergence of the evaluation time per candidate.
The data acquisition time is small due to our implementation of restless measurements \cite{Rol2017} which allow for a $100~\rm{kHz}$ repetition rate.
The data analysis time is negligible in comparison to the three main contributions and is not included in the analysis. 

\subsection{Numerical results}

To obtain the numerical results in Fig.~\ref{fig:results} (blue dashed and red dotted lines) we model the qubit as a driven an-harmonic oscillator with a Hilbert space of dimension $d=4$. We simulate the quantum dynamics and perform quantum optimal control with the q-optimize package \cite{Machnes2020} which is built using TensorFlow \cite{tensorflow}. This allows for fast simulations of the dynamics of open and closed quantum systems and gradient-based optimization of the goal function via automatic-differentiation. 

For each of the $\pm X/2$ and $\pm Y/2$ gates, we convolve both the DRAG and the piecewise-constant pulses, as sampled by the AWG, with a Gaussian window to produce a $0.3~\rm{ns}$ rise-time, thus, emulating the limited bandwidth of the AWG.
We obtain the dynamics as superoperator matrices $\Lambda_{\pm X/2}$ and $\Lambda_{\pm Y/2}$ by solving the master equation in Lindblad form, which includes $T_1$ and $T_2$. 
Next, we compose the gates $C_i$ in the Clifford group $C$ from these atomic operations, for example $C_6=\Lambda_{-X/2}\circ \Lambda_{-Y/2}\circ \Lambda_{X/2}$.

The pulse parameters $\mathcal{S}$ are optimized using the L-BFGS gradient search \cite{L-BFGS} to maximize the average Clifford gate fidelity $F_C=\left\vert{C}\right\vert^{-1}\sum_i F_{C_i}$. This optimizer is more efficient for numerical simulations than the CMA-ES optimizer if measurement noise is neglected and gradients can be computed. 
The fidelity of each Clifford gate is $F_{C_i}= \frac{1}{d + 1}\left(d\mathcal{X}_{0,0}  + 1\right)$. Here, $\mathcal{X}_{0,0}$ is the $\left(0,0\right)^\text{th}$ element of the Choi matrix $\mathcal{X}$ representing the gate error $\Lambda=\widetilde{C}_i^\dagger \circ C_i$ between the perfect Clifford gate $\widetilde{C}_i$ and the implemented Clifford gate $C_i$ \cite{Magesan2011_2}.
In addition to $T_1$ and $T_2$ relaxation we include an amplitude-dependent dephasing error channel  $D(\rho) = \left(1-\gamma_\phi\right) \mathcal{I} \rho \mathcal{I}  + \gamma_\phi Z \rho Z$, applied as a superoperator to each single-qubit operation composing a Clifford gate, e.g. $C_6=D \circ \Lambda_{-X/2}\circ D \circ \Lambda_{-Y/2}\circ D\circ \Lambda_{X/2}$. The dephasing strength is $\gamma_\phi=k\cdot\Omega\cdot  t_\text{gate}$, where $\Omega$ and $t_\text{gate}$ are the average amplitude and length, respectively, of the pulses implementing the gates. The constant $k$ is chosen to match the data.

\section{Acknowledgements}

We thank A. Fuhrer, M. Ganzhorn, M. Mergenthaler, P. Mueller, S. Paredes, M. Pechal, and G. Salis for insightful discussions as well as R. Heller and H. Steinauer for technical support. We also acknowledge useful discussions and the provision of qubit devices with the quantum team at IBM T. J. Watson Research Center, Yorktown Heights. This work was supported by the European Commission Marie Curie ETN QuSCo (Grant Nr. 765267), the IARPA LogiQ program under contract W911NF-16-1-0114-FE and the ARO under contract W911NF-14-1-0124. S.M. and F.W. acknowledge support from the the project OpenSuperQ (Grant Nr. 820363) and project VERTICONS (BMBF, Project 13N14872)

\section{Author Contributions}

All authors developed the idea for the experiment; M. W. performed the measurements and analysed the data;
M.W. and D.E. implemented the control methods. 
F.R. and S.M. carried out the numerical simulations. M.W., D.E., F.R. and S.F. wrote the manuscript.
All authors contributed to the discussions and interpretations of the results.

\end{document}